\documentclass[preprint,10pt]{elsarticle}

\usepackage{amssymb}
\usepackage{amsmath}
\usepackage[table]{xcolor}
\usepackage{lineno}
\usepackage{comment}
\usepackage{bm}
\usepackage{caption}
\usepackage{geometry}
\usepackage{booktabs}
\usepackage{multirow}
\usepackage{color}
\usepackage[colorlinks,linkcolor=blue]{hyperref}
\usepackage[nameinlink]{cleveref}

\crefname{figure}{Fig.}{Figs.}
\crefformat{equation}{Eq.~#2(#1)#3}
\crefformat{section}{Section~#2#1#3}
\AtBeginDocument{%
\let\citet\cite
}

\journal{Elsevier}
\begin{document}
\captionsetup[figure]{name={\textbf{Fig.}},labelsep=period}
\captionsetup[table]{name={\textbf{Table}},labelsep=period}
\begin{frontmatter}



\title{Energy-based physics-informed neural network for frictionless contact problems under large deformation }


\author[inst1,inst2]{Jinshuai Bai}
\author[inst1]{Zhongya Lin}
\author[inst3,inst5]{Yizheng Wang}
\author[inst2,inst4]{Jiancong Wen}
\author[inst3]{Yinghua Liu}
\author[inst5]{Timon Rabczuk}
\author[inst2]{YuanTong Gu\corref{cor1}}
\author[inst1]{Xi-Qiao Feng\corref{cor1}}
\cortext[cor1]{Corresponding authors: yuantong.gu@qut.edu.au (YuanTong Gu), fengxq@tsinghua.edu.cn (Xi-Qiao Feng)}

\affiliation[inst1]{organization={Institute of Biomechanics and Medical Engineering, AML, Department of Engineering Mechanics, Tsinghua University},
            city={Beijing},
            postcode={100084}, 
            country={China}}

\affiliation[inst2]{organization={School of Mechanical, Medical and Process Engineering, Queensland University of Technology},
            city={Brisbane},
            postcode={4000}, 
            state={Queensland},
            country={Australia}}

\affiliation[inst3]{organization={Department of Engineering Mechanics, AML, Tsinghua University},
            city={Beijing},
            postcode={100894}, 
            country={China}}

\affiliation[inst4]{organization={Institute of Aeronautics and Astronautics, Nanchang University},
            city={Nanchang},
            postcode={330031}, 
            country={China}}

\affiliation[inst5]{organization={Institute of Structural Mechanics, Bauhaus Universität Weimar},
            city={Weimar},
            postcode={99423}, 
            state={Thuringia},
            country={Germany}}

\begin{abstract}
Numerical methods for contact mechanics are of great importance in engineering applications, enabling the prediction and analysis of complex surface interactions under various conditions. In this work, we propose an energy-based physics-informed neural network (PINN) framework for solving frictionless contact problems under large deformation. Inspired by microscopic Lennard-Jones potential, a surface contact energy is used to describe the contact phenomena. To ensure the robustness of the proposed PINN framework, relaxation, gradual loading and output scaling techniques are introduced. In the numerical examples, the well-known Hertz contact benchmark problem is conducted, demonstrating the effectiveness and robustness of the proposed PINN framework. Moreover, challenging contact problems with the consideration of geometrical and material nonlinearities are tested. It has been shown that the proposed PINN framework provides a reliable and powerful tool for nonlinear contact mechanics. More importantly, the proposed PINN framework exhibits competitive computational efficiency to the commercial FEM software when dealing with those complex contact problems. The codes used in this manuscript are available at \href{https://github.com/JinshuaiBai/energy\_PIsNN\_Contact}{https://github.com/JinshuaiBai/energy\_PINN\_Contact}.

\end{abstract}


\end{frontmatter}


\section{Introduction}
\label{sec:sample1}
Contact and interaction between objects are ubiquitous in nature and industrial production. Accurately simulating the deformation of bodies during contact is crucial for understanding and predicting mechanical behaviour. Contact problems are inherently nonlinear and are a fundamental aspect of solid mechanics  \cite{wriggers_Contact_textbook}. Over the years, significant efforts have been dedicated to addressing these problems, with mesh-based methods emerging as a predominant approach \cite{Sun2023_cont_SFEM,YUE2018110,Liu2022_contact,Wriggers2013Third}. Additionally, meshless methods, employing both weak and strong formulations, have been proposed to tackle challenges in contact mechanics \cite{Li2007_cont,BELAASILIA2018103,ALMASI2019112597,ALMASI2023104291,Beel2023}. 

The complexities of contact mechanics, such as contact discretization and boundary smoothness, make it a particularly challenging area of study \cite{laursen_Contact_textbook}. To overcome these challenges, isogeometric analysis (IGA) frameworks based on non-uniform rational B-splines (NURBS) have been developed. These frameworks enable exact representations of complex geometries and provide superior continuity properties \cite{temizer2012three, de2012mortar,temizer2011contact, lu2011isogeometric, de2011large}. Beyond issues of contact smoothness and discretization, the enforcement of inequality constraints represents another critical aspect of contact mechanics. Various numerical techniques, such as the Lagrange multiplier method, the augmented Lagrangian formulation, and mortar methods, have been extensively developed to address this issue \cite{FERNANDEZ2020113288, Bozorgmehri2021}.

In recent years, a novel kind of deep learning (DL) framework \cite{lecun2015deep, liu2022machine}, namely the physics-informed neural networks (PINNs), has merged as a powerful and promising tool for solving partial differential equations \cite{raissi2018deep, raissi2019physics, Cai2021, wang2024artificial}, and therefore attracted significant attention in the computational mechanics \cite{HFM_Science, WESSELS2020113127, BAI2022114740, Li2024}. Numerous numbers of PINN-based computational solid mechanics frameworks have been proposed \cite{Bai2023_LSWR, haghighat2021physics, samaniego2020energy, CENN_WYZ}. Among those frameworks, neural networks regulated by the energy form loss provide the most robust results with favourable efficiency. Those energy-based PINN frameworks are also known as the deep energy method (DEM) \cite{samaniego2020energy}. Great performances of those PINN frameworks have been witnessed in a variety of applications, including linear elasticity \cite{LIWEI_energy_PINN, samaniego2020energy}, hyperelasticity \cite{NGUYENTHANH_DEM_Hyper}, elastoplasticity \cite{NIU2023105177,HE2023103531}, fracture \cite{GOSWAMI2019106608,GOSWAMI2020102447,zhao2024dedem}, large deformation problems \cite{BaiRPIMNNS}, and topology optimisation \cite{ZHANG2021114083, JEONG2023115484, JEONG2023116401, jeong2025advanced}. Besides, adaptive sampling strategies \cite{tang2023pinns, mao2023physics} have been proposed to avoid the overfitting issue and enhance the accuracy of predictions. In addition, traditional PDE solving techniques have been introduced to the energy-based PINN frameworks, ending up with establishing novel methods for solid mechanics challenges. For example, complementary energy form and boundary integration form lead to novel deep complementary energy method \cite{wang2023dcem} and boundary-informed neural network frameworks \cite{BINN_WYZ}, respectively. Novel neural network structures are also implemented. Bai et al. \cite{BaiRPIMNNS} applied radial basis networks to deal with problems involving both material and geometrical nonlinearity. It has been demonstrated that such a framework can easily capture instability phenomena and is locking-free when modelling nearly incompressible materials. Wang et al. \cite{wang2024kolmogorov} implement the Kolmogorov Arnold network (KAN) with physics knowledge. By doing so, the KAN shows exceptional performances for multi-scale and material discontinuity problems.

Despite its notable success in computational mechanics, only limited exploration regarding PINNs has been conducted on contact problems. Sahin et al. \cite{Sahin2024} proposed a PINN framework under the strong form formulation with the classical Karush–Kuhn–Tucker constraint to solve the forward and inverse contact problems.  Efforts have also been directed toward addressing interfacial discontinuities arising from the consideration of multi-material systems \cite{DIAO2023116120}. However, in those existing and limited literature, either small deformation problems are discussed or the contact surfaces remain unchanged throughout the training, e.g., neither new contact area generated nor sliding between contact surfaces is considered.  It has been demonstrated that the PINN framework is very effective for material nonlinearities and large deformation problems \cite{NGUYENTHANH_DEM_Hyper, BaiRPIMNNS}. Besides, the energy-based PINNs are computationally more efficient and robust than the vanilla PINNs \cite{CENN_WYZ}. Therefore, it is of great interest to implement energy-based PINNs to solve contact problems. Moreover, the effectiveness and performance of the energy-based PINNs for contact problems with complex nonlinearities are also worth investigating. 

To address this research gap, in this work, we propose an energy-based PINN framework for contact problems with both material and geometrical nonlinearities. The contact behaviour is described by a contact potential and together optimised by gradient descendant algorithms. Moreover, additional numerical treatments, including the relaxation, gradual loading and output scaling are also introduced to improve the robustness of the proposed framework. The Hertz contact benchmark problems under small deformation assumption are examined to show the performances of the proposed framework. More importantly, various contact examples considering both the material and the geometrical nonlinearities are tested, including the rubber ironing, the rubber ring contact instability and the compression of two rings. The deformed configurations and stress contours demonstrate the effectiveness and substantial potential of the proposed PINN framework for complex nonlinear mechanics. Notably, the proposed framework is very straightforward to implement numerically and can easily incorporate experimental data. Additionally, the PINN framework demonstrates competitive computational efficiency compared to commercial FEM software when addressing complex contact problems.

The remainder of the paper is organised as follows: In \Cref{sec:sec2}, the basic conceptions of energy-based PINN for solid mechanics are briefly introduced. In \Cref{sec:sec3},  the frictionless contact achieved by using the surface potential and its implementation based on the PINN framework are proposed. The Hertz contact benchmark problem is also examined and discussed in detail. In \Cref{sec:sec4}, numerical examples involving large deformation and material nonlinearities are presented, including the rubber ironing, the rubber ring contact instability and the compression of two rings. In  \Cref{sec:sec5}, the conclusions of this work are summarised. 

\section{Energy-based PINN for nonlinear solid mechanics}
\label{sec:sec2}

\subsection{Loss function informed by potential energy}

We first recap the basic conceptions of the energy-based PINN in solid mechanics. In this method, neural networks are used to approximate the admissible displacement solution $ {\boldsymbol{u}}(\boldsymbol{x}) $ satisfying the essential boundary condition:

\begin{equation}
\begin{aligned}
\boldsymbol{u}(\boldsymbol{x})&\approx\boldsymbol{F}(\boldsymbol{x};\boldsymbol{\theta})\\
\boldsymbol{u}(\boldsymbol{x})&=\bar{\boldsymbol{u}}(\boldsymbol{x}), \qquad \boldsymbol{x}\in\Gamma^{\boldsymbol{u}}
\label{eq:Neural_net_mapping}
\end{aligned}
\end{equation}

\noindent where $\boldsymbol{F}(\cdot)$ denotes a neural network mapping and $\bm{\theta}$ denotes the trainable parameters in the network. $\bar{\boldsymbol{u}}(\boldsymbol{x})$ is the prescribed essential boundary conditions on essential boundary $\Gamma^{\boldsymbol{u}}$. Note that the neural network structure has been introduced in many PINN-based computational mechanics literature, and different types of neural networks can be found in \cite{ZHUANG2021104225,BAI2023116290,ZHOU2024112695}. Referring to the principle of minimum potential energy, the solution to the solid mechanics problem can be obtained by finding a $ \boldsymbol{u}(\boldsymbol{x}) $ that minimises the potential energy of the solid system

\begin{equation}
\boldsymbol{u}(\boldsymbol{x})=\arg \min_{\substack{
   \boldsymbol{u}(\boldsymbol{x})
  }} \Pi(\boldsymbol{u}(\boldsymbol{x})),
\end{equation}

\noindent where $\Pi(\cdot) $ is the overall potential energy functional and can be calculated by

\begin{equation}
\Pi=E_{\text{in}}-E_{\text{ex}},
\label{eq:overall_energy}
\end{equation}

\noindent where $ E_{\text{in}} $ and $ E_{\text{ex}} $ are the strain energy and the potential energy of external forces, respectively. For elastic material, the strain energy can be calculated by

\begin{equation}
E_{\text{in}}=\int_{\Omega} \Psi(\boldsymbol{F}) \mathrm{d}\Omega,
\end{equation}

\noindent where $ \Psi(\cdot) $ and $ \boldsymbol{F} $ are the strain energy density and deformation gradient tensor, respectively \cite{SCHAEFFERKOETTER2024117307}. The potential energy of external forces can be obtained by

\begin{equation}
E_{\text{ex}}=\int_{\Gamma^{\boldsymbol{t}}} \boldsymbol{\bar{t}} \cdot \boldsymbol{{u}} \mathrm{d}\Gamma,
\end{equation}

\noindent where $\bar{\boldsymbol{t}}$ is the given traction force on the natural boundary $\Gamma^{\boldsymbol{t}}$. Many weak-form computational mechanics methods are established based on the principle of minimum potential energy, such as FEM, element free Galerkin (EFG) method \cite{zhu_EFG} and point interpolation method (PIM) \cite{GRLIU_PIM}. They are considered to be more stable than those  based on the strong form formulation, and PINN-based computational mechanics frameworks are no exception \cite{bai2023introduction}. It has been reported that energy-based PINNs are computationally more efficient and better at capturing the stress concentration \cite{LIWEI_energy_PINN,bai2023introduction}.

\subsection{Boundary condition imposition}
As mentioned above, the natural (traction) boundary conditions are already embedded in the potential energy. Auxiliary techniques are required to impose the essential boundary conditions. Consider the essential boundary condition is defined on $\Gamma^{\boldsymbol{u}}$
\begin{equation}
  \boldsymbol{u}(\boldsymbol{x})=\bar{\boldsymbol{u}}(\boldsymbol{x}), \qquad \boldsymbol{x}\in\Gamma^{\boldsymbol{u}} 
\end{equation}
\noindent where $\bar{\boldsymbol{u}}(\boldsymbol{x})$ is the prescribed essential boundary conditions on $\Gamma^{\boldsymbol{u}}$. Currently, the essential boundary conditions in energy-based PINNs can be imposed in soft or hard manners. 

The so-called "\textit{soft}" boundary condition imposition technique in PINNs is also known as the penalty method \cite{Barrett1986}. It produces solutions that approximately satisfy essential boundary conditions. To implement it, a least-square functional, $\Pi_{\text{EBC}}$, is first formulated
\begin{equation}
\Pi_{\text{EBC}}=\int_{\Gamma^{u}} \Vert\boldsymbol{u}(\boldsymbol{x})-\bar{\boldsymbol{u}}(\boldsymbol{x})\Vert_{2}^{2} \mathrm{d}\Gamma = \int_{\Gamma^{u}} \left( [\boldsymbol{u}(\boldsymbol{x})-\bar{\boldsymbol{u}}(\boldsymbol{x})]\cdot[\boldsymbol{u}(\boldsymbol{x})-\bar{\boldsymbol{u}}(\boldsymbol{x})]\right)\mathrm{d}\Gamma,
\end{equation}
where $\Vert\cdot\Vert_{2}$ is the 2-norm on the essential boundary. As observed, $\Pi_{\text{EBC}}$ only reaches its minimal value when the predicted displacements exactly equal the given boundary conditions. It is then added to the overall potential energy functional with a penalty factor $\kappa$
\begin{equation}
    \Pi^{*}=\Pi+\kappa\Pi_{\text{EBC}}.
    \label{eq:penalty_method_functional}
\end{equation}
\noindent where $\Pi^{*}$ is the modified energy functional. Referring to the penalty method, $\kappa$ should be a positive factor. With increasing $\kappa$, the solution obtained by \Cref{eq:penalty_method_functional} will recover to the solution of the original problem. In numerical implementation, residuals are inevitable on the essential boundaries, such penalty method is therefore called the soft boundary condition imposition technique. 

To impose the essential boundary conditions in the "hard" way, the neural network output is tailored so that it can naturally satisfy the essential boundary conditions
\begin{equation}
    \boldsymbol{u}(\boldsymbol{x})=\boldsymbol{F}(\boldsymbol{x})\odot \boldsymbol{g}(\boldsymbol{x})+\bar{\boldsymbol{u}}(\boldsymbol{x}),
    \label{eq:Modified_NN_Map}
\end{equation}
where $\odot$ is the element-wise product, and the distance network $\boldsymbol{g}(\boldsymbol{x})$ represents the shortest distance from a given point $\boldsymbol{x}$ to the essential boundary:
\begin{equation}
\boldsymbol{g}(\boldsymbol{x}) = \min_{\boldsymbol{y} \in \Gamma^{u}}\sqrt{\Vert\boldsymbol{x}-\boldsymbol{y}\Vert_{2}^{2}}=\min_{\boldsymbol{y} \in \Gamma^{u}} \sqrt{(\boldsymbol{x} - \boldsymbol{y}) \cdot (\boldsymbol{x} - \boldsymbol{y})}.
\end{equation}

There are two main advantages of using the hard boundary condition imposition technique. First, the essential boundary condition can be exactly imposed. Second, the physics-informed loss function can include fewer loss terms. The neural network training can be regarded as a multi-task learning process if multiple loss terms exist in the final loss function. Using too many loss terms may induce the imbalance training issue, and thus hyperparameters are required to balance the residuals from different loss terms \cite{Bai2023_LSWR}. Tuning hyperparameters before each loss term can be annoying. For simple geometries, the explicit form of $\boldsymbol{g}(\boldsymbol{x})$ can be obtained. For complex boundary geometries, the distance function can be constructed in replacement of $\boldsymbol{g}(\boldsymbol{x})$.  For more details, please refer to \cite{BC_imoposition_SUKUMAR2022114333}. 

In this study, the hard boundary condition enforcement technique is directly utilised to fixed boundary conditions. For displacement loading, the soft boundary condition enforcement is initially applied, followed by the use of the hard boundary condition technique.

\subsection{Neural network training and its pseudo-dynamic nature}
\label{sec:2:3}
\begin{figure}
    \centering
    \includegraphics[width=1\linewidth]{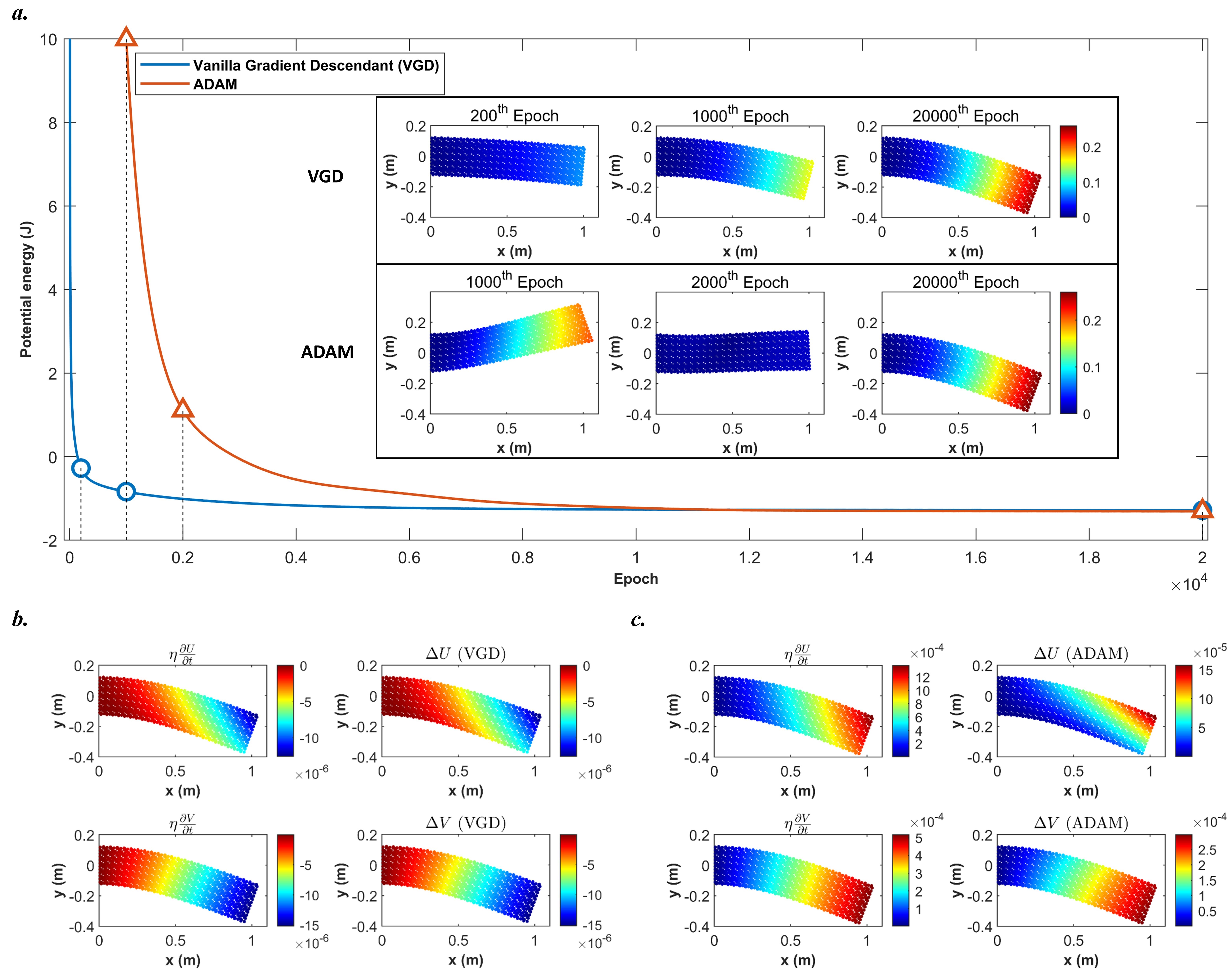}
    \caption{(a) The training dynamics of energy-based PINNs for solving a cantilever beam ($0.25$ m $\times1$ m) problem with a linear elastic material. The Young's modulus and the Poisson's ratio are $1\times10^4$ Pa and $0.3$, respectively. The left boundary of the beam is fixed. A parabolic distributed force of  $10$ N is downwardly applied on the right boundary. Two feedforward neural networks with $3$ hidden layers and $5$ neurons per layer are used for predicting displacements. When using gradient descendant algorithms, the overall potential energy decreases during the training of energy-based PINNs. The intermediate absolute displacement mappings are shown. Along with the decreasing energy functional, the beam seems to be dynamically bent to the equilibrium state. Note that the VGD in the figure refers to the vanilla gradient descendant algorithm. A learning rate of $1\times10^{-4}$ is applied. (b) Comparisons between $\eta\frac {\partial \boldsymbol{u}}{\partial t}$ and actual displacement increment when using the VGD at the $20000^{\text{th}}$ epochs. (c) Comparisons between $\eta\frac {\partial \boldsymbol{u}}{\partial t}$ and actual displacement increment when using the ADAM optimiser at the $20000^{\text{th}}$ epochs. }
    \label{fig:dynamical_bent}
\end{figure}
To solve the problem, optimisers are applied to minimise the loss function by iteratively adjusting those trainable parameters, $\boldsymbol{\theta}$. Among all, gradient descendant optimisers have earned great favour due to their effectiveness and efficiency. Currently, the ADAM optimiser is the most popular optimiser in PINN training \cite{kingma2014adam}.  In the energy-based PINNs, the energy functional at a given set of displacement mappings is treated as the loss function. Therefore, the training of the energy-based PINN can be regarded as a way to dissipate the overall potential energy. In other words, the energy-based PINNs are trained in a pseudo-dynamic manner. This can be verified by the example shown in  \Cref{fig:dynamical_bent} (a), where both the vanilla gradient descendant (VGD) algorithm and the ADAM optimiser are applied to train neural networks for a 2D cantilever beam problem. After initialisation, neural networks map to  random displacement fields. During the training, with the decreasing of the potential energy functional, the intermediate displacement mappings are shown. Along with the decreasing energy functional, the beam seems to be dynamically bent to the equilibrium state. Eventually, the functional loss stably converges around $2\times10^4$ epochs. Meanwhile, it can be noticed that the bending rates of the beam vary from the use of different optimisers. The ADAM optimiser incorporates the momentum of the gradient information. Hence, it can adjust the learning rate of each trainable parameter. By so doing, the use of the ADAM optimiser can significantly avoid loss oscillations during training and enhance the training robustness.  

If gradient descendant algorithms are selected, the pseudo-velocity of the displacement mappings by neural networks during training can be approximately calculated by
\begin{equation}
\frac {\partial u}{\partial t}\approx- \frac{\partial F(\boldsymbol{x};\boldsymbol{\theta})^{\text{T}}}{\partial \boldsymbol{\theta}}\frac{\partial \Pi}{\partial \boldsymbol{\theta}},
\label{eq:pseudo-veloctiy}
\end{equation}
where $F(\boldsymbol{x};\boldsymbol{\theta})$ is the neural network displacement mapping, as presented in \Cref{eq:Neural_net_mapping}. Moreover, the increment of displacement can obtained by 
\begin{equation}
\Delta u \approx \eta \frac{\partial u}{\partial t}.
\end{equation}
The derivation is given in \ref{sec:sample:appendix}. These equations can be further verified by \Cref{fig:dynamical_bent} (b) and (c). As observed, the pseudo-velocity can effectively predict the displacement increment during training when the VGD algorithm is selected. Although small departures can be observed when using the ADAM optimiser, the overall pseudo-velocity is quite similar to the displacement increment after the energy loss is stabilised. It should be noted that the pseudo-dynamics nature exhibited by neural network training is not identical to second-order ordinary differential systems. In addition, the VGD converges faster at the beginning of the training. However, for complex nonlinear problems, the VGD may trapped in local optima, while the ADAM optimiser can provide more reliable predictions (see \ref{sec:app_gd_nolin}). We highlight that the pseudo-dynamics nature of the energy-based PINN can be utilised as a gradual loading process till the equilibrium state, equivalent to the iterative solvers in the traditional methods when dealing with nonlinear problems. Besides, the pseudo-velocity can help properly select the learning rate so that no penetration will happen during the training process. Those numerical techniques are explained in the following section. 

\section{Frictionless contact potentials and numerical implementations}
\label{sec:sec3}
\subsection{Frictionless surface contact potential}
\begin{figure}
\centering
\includegraphics[scale=1]{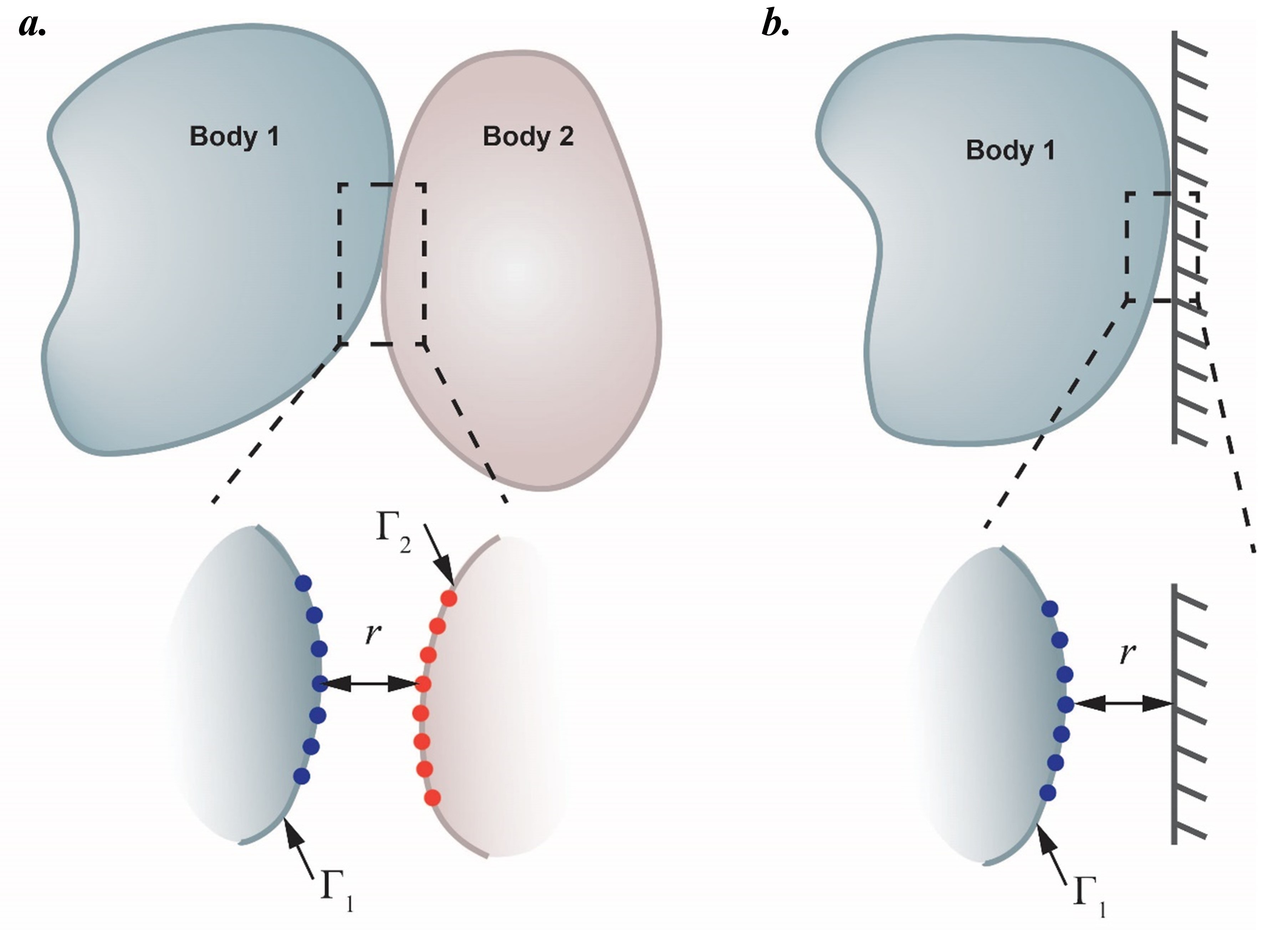}
\caption{Schematics of contact model discretisation \cite{wriggers_Contact_textbook}. (a) Point-to-point contact model. (b) Point-to-surface model.}
\label{fig:Types_of_model}
\end{figure}

\begin{figure}
    \centering
    \includegraphics[width=0.9\linewidth]{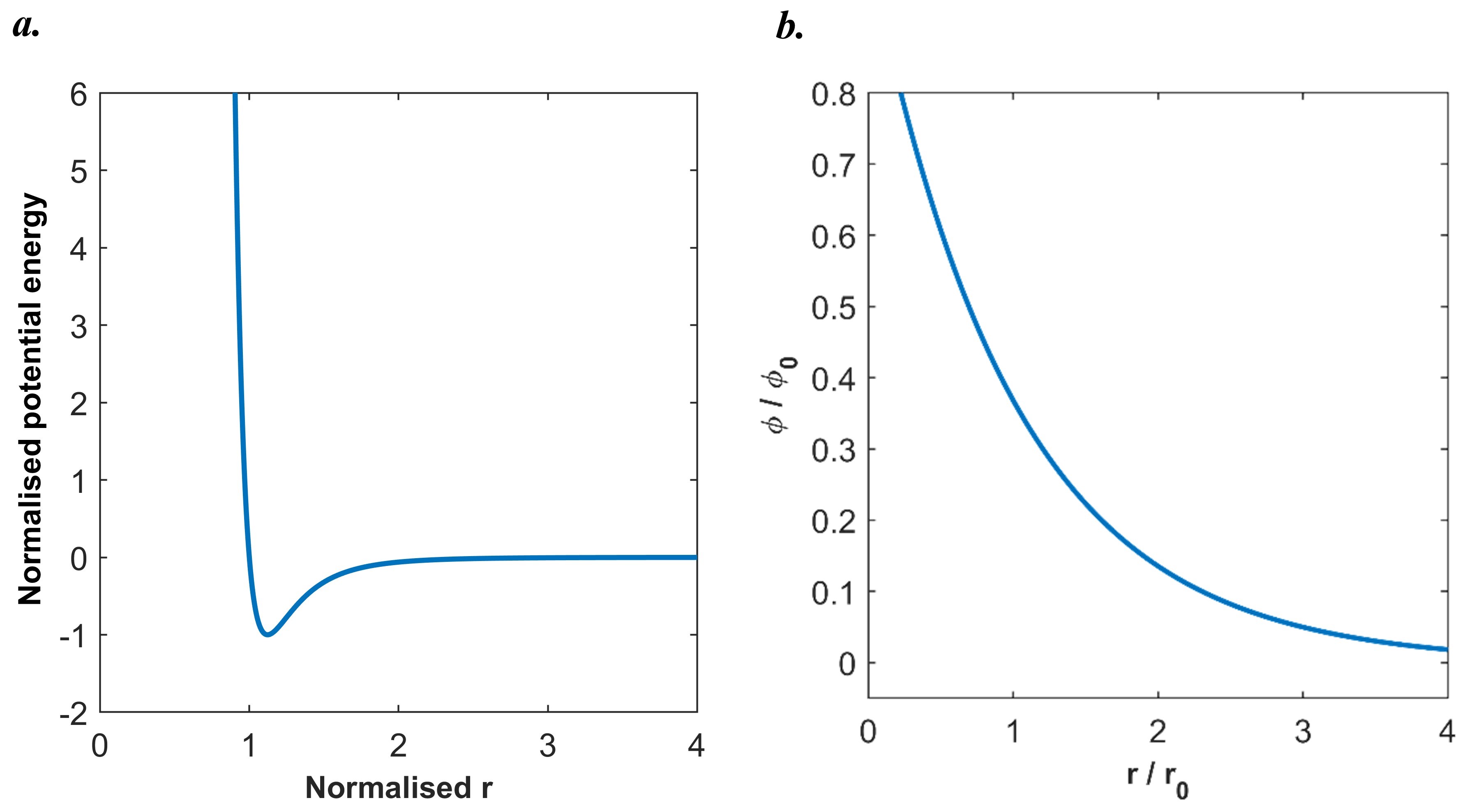}
    \caption{(a) The Lennard-Jones potential plot; (b) The exponential form surface contact potential used in this work. Note that the $\phi_0$ and $r_0$ are the pre-defined potential constant and the effective radius of the discrete points, respectively. }
    \label{fig:potentialplots}
\end{figure}

In general contact problems, consider two bodies are getting close to each other, as shown in  \Cref{fig:Types_of_model} (a). From the microscopic point of view, it is the molecular and atom forces between two bodies that prevent them from penetrating each other. Such repulsive phenomenon is usually described by potentials, for example, the well-known Lennard-Jones (LJ) potential \cite{LJ_ref}, as presented in \Cref{fig:potentialplots} (a).  Therefore, bringing this idea to the macroscale, a penalty-liked surface contact potential can be used to mimic the microscopic LJ potential between molecules and atoms. In this case, the overall potential energy of the conservation system can be re-written as
\begin{equation}
\Pi=E_{\text{in}}-E_{\text{ex}}+E_{\text{c}},
\end{equation}
\noindent where $E_{\text{c}}$ is the surface contact potential energy and can be obtained by \cite{SAUER2013369} 
\begin{equation}
E_{\text{c}}=\int_{\Gamma_1} \int_{\Gamma_2} \beta_1 \beta_2 \phi (r) \mathrm{d}\Gamma_1 \mathrm{d}\Gamma_2,
\label{eq:potential_eq}
\end{equation}
\noindent where $\phi$ and $r$ are the surface contact potential and the relative distance between contact surfaces, respectively. $\beta$ is the surface density. The subscribes 1 and 2 denote the two contact surfaces. As observed,  surface integrations are done on both surfaces of bodies. It is clear to find that the LJ potential is not a monotonic function, where it goes down first and then increases sharply when $r\to0$. Nevertheless, given that this non-monotonic area only appears when the distance between the atom and molecular is close enough, which generally does not appear in the macroscopic contact problems. Therefore, a monotonic function can be applied as the surface contact potential, as long as the macroscopic potential increases  steeply when $r\to0$ and decays fastly when $r\to\infty$. Here, an exponential form potential is applied in this work and defined as
\begin{equation}
\phi(r)=\phi_0 \exp{\left(- \frac{r}{r_0}\right)}, \label{contact potential}
\end{equation}
\noindent where $\phi_0$ and $r_0$ are the pre-defined potential constant and the effective radius of the discrete points, respectively. The plot of the repulsive potential is given in \Cref{fig:potentialplots} (b). Therefore, the corresponding contact force can be calculated by
\begin{equation}
    \begin{aligned}
\boldsymbol{F}_{\text{c}} &= \int_{\Gamma_1} \int_{\Gamma_2} \beta_1 \beta_2 \frac{\partial \phi(r)}{\partial \boldsymbol{r}} \mathrm{d}\Gamma_1 \mathrm{d}\Gamma_2 \\ &= \int_{\Gamma_1} \int_{\Gamma_2} -\beta_1 \beta_2 \frac{\phi_0}{r_0} \frac{\boldsymbol{r}}{r} \exp{\left(-\frac{r}{r_0}\right)}\mathrm{d}\Gamma_1 \mathrm{d}\Gamma_2.
\label{eq:contact_force_eq}
    \end{aligned}
\end{equation}
\noindent It is worth noting that the idea of using the surface potential for contact problems has been implemented in \cite{SAUER2013369}. Besides, the rigorous derivations regarding the contact forces and variational formulation have been discussed in detail. However, in the framework of the proposed energy-based PINN, the potential energy functional is directly used and is minimised by powerful neural network optimisers. Meanwhile, we note that contact potentials can be in different forms and not necessarily identical to \Cref{contact potential}. Other contact potentials can be selected according to different contact problems. More contact potentials can be found in \cite{SAUER2013369}. This work will only focus on demonstrating the effectiveness of the proposed PINN framework for contact problems. Therefore, only one contact potential is tested in the following numerical examples. 

 To numerically obtain the surface contact potential energy, sample points are placed to the possible contact area on both surfaces of the two bodies. In this work, this kind of contact model is called the point-to-point (PP) model, as shown in \Cref{fig:Types_of_model} (a). Such kind of contact model can generally apply to different irregular geometries. 

For some cases where regular geometry is used, as shown in \Cref{fig:Types_of_model} (b). In these scenarios, only discrete sample points are placed on the irregular surface. In this work, we call this kind of contact model the point-to-surface (PS) model. Therefore, the surface contact potential energy of the PS contact model can be simplified as 
\begin{equation}
E_{\text{c}}=\int_{\Gamma_1} \beta_1 \phi (r) \mathrm{d}\Gamma_1 ,
\end{equation}
where $r$ is always perpendicular to the surface of the regular surface. Such a PS model is less generalised but can greatly improve both computational efficiency and accuracy. In the following numerical examples, both the PP and the PS models are applied to deal with different contact scenarios. More detailed discussions between those two contact models are available in  \Cref{sec:3:4} .

It is worth noting that, such contact potential can also be used to describe the frictional contact phenomenon. However, in this work, we only use this potential to deal with frictionless contact problems. The friction force is a non-conservative force. Hence, frictional contact problems will be covered in our future work.

\subsection{Numerical implementation}
\begin{figure}
\centering
\includegraphics[scale=0.9]{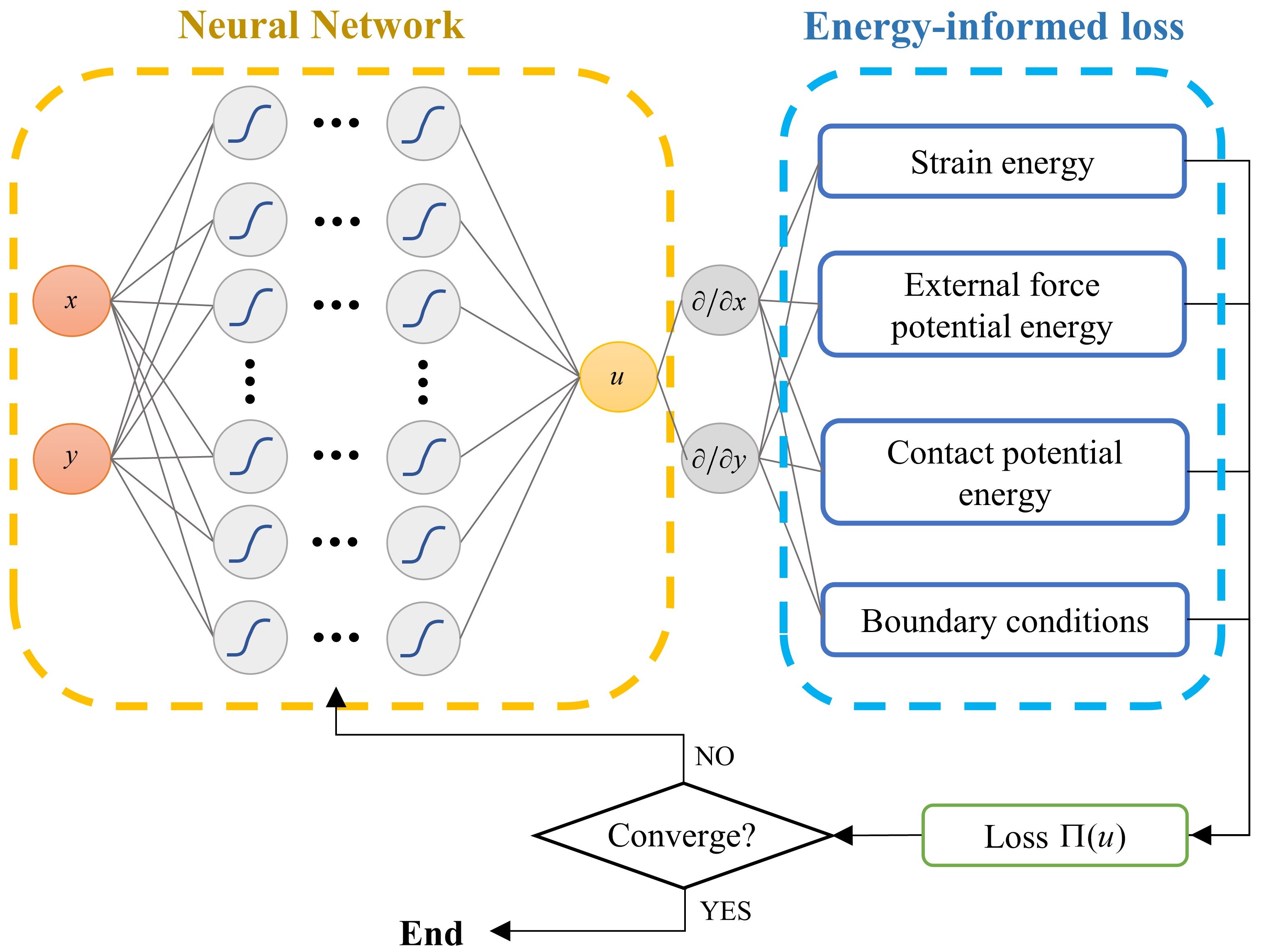}
\caption{The schematic of the proposed energy-based PINN framework for contact problems.}
\label{fig:Fig.2}
\end{figure}

As shown in \Cref{contact potential}, the distances between the contact boundary nodes are necessary. Here, all the formulas are based on 2D problems and they can be easily extended to 3D scenarios. Consider the numbers of contact points on the potential contact surfaces are $n$ and $m$, respectively. The relative distance between contact points can be stored in a matrix of size $n\times m$ 

\begin{equation}
r_{n\times m}=\sqrt{(\Delta \boldsymbol{u})^2 + (\Delta \boldsymbol{v})^2},
\end{equation}

\noindent where

\begin{equation}
\begin{aligned}
\Delta \boldsymbol{u}_{n\times m} = \boldsymbol{x}_{n\times 1} \otimes \boldsymbol{1}_{m\times 1} - (\boldsymbol{x}_{m\times 1} \otimes \boldsymbol{1}_{n\times 1})^{\text{T}}\\
\Delta \boldsymbol{v}_{n\times m} = \boldsymbol{y}_{n\times 1} \otimes \boldsymbol{1}_{m\times 1} - (\boldsymbol{y}_{m\times 1} \otimes \boldsymbol{1}_{n\times 1})^{\text{T}}\\
\end{aligned}
\end{equation}

\noindent where $\boldsymbol{1}$ is the vector of ones, and the notation $\otimes$ denotes the outer product. In Python, the outer product can be directly used as the multiplication by TensorFlow2. Therefore, the potential between each pair of points can be calculated as 
\begin{equation}
    \phi(r)_{n\times m}=\phi_0 \exp{\left(-\frac{r_{n\times m}}{r_0}\right)}.
\end{equation}
This equation is then substituted into \Cref{eq:potential_eq} to calculate the overall contact potential. We note that the above implementation is compatible with the automatic differentiation technique. Thus, the contact energy added to the final loss can be directly minimised by neural network optimisers. The overall PINN framework proposed for contact problems is shown in \Cref{fig:Fig.2}. 

\subsection{Relaxation, gradual loading and output scaling}
\begin{figure}
    \centering
    \includegraphics[width=1\linewidth]{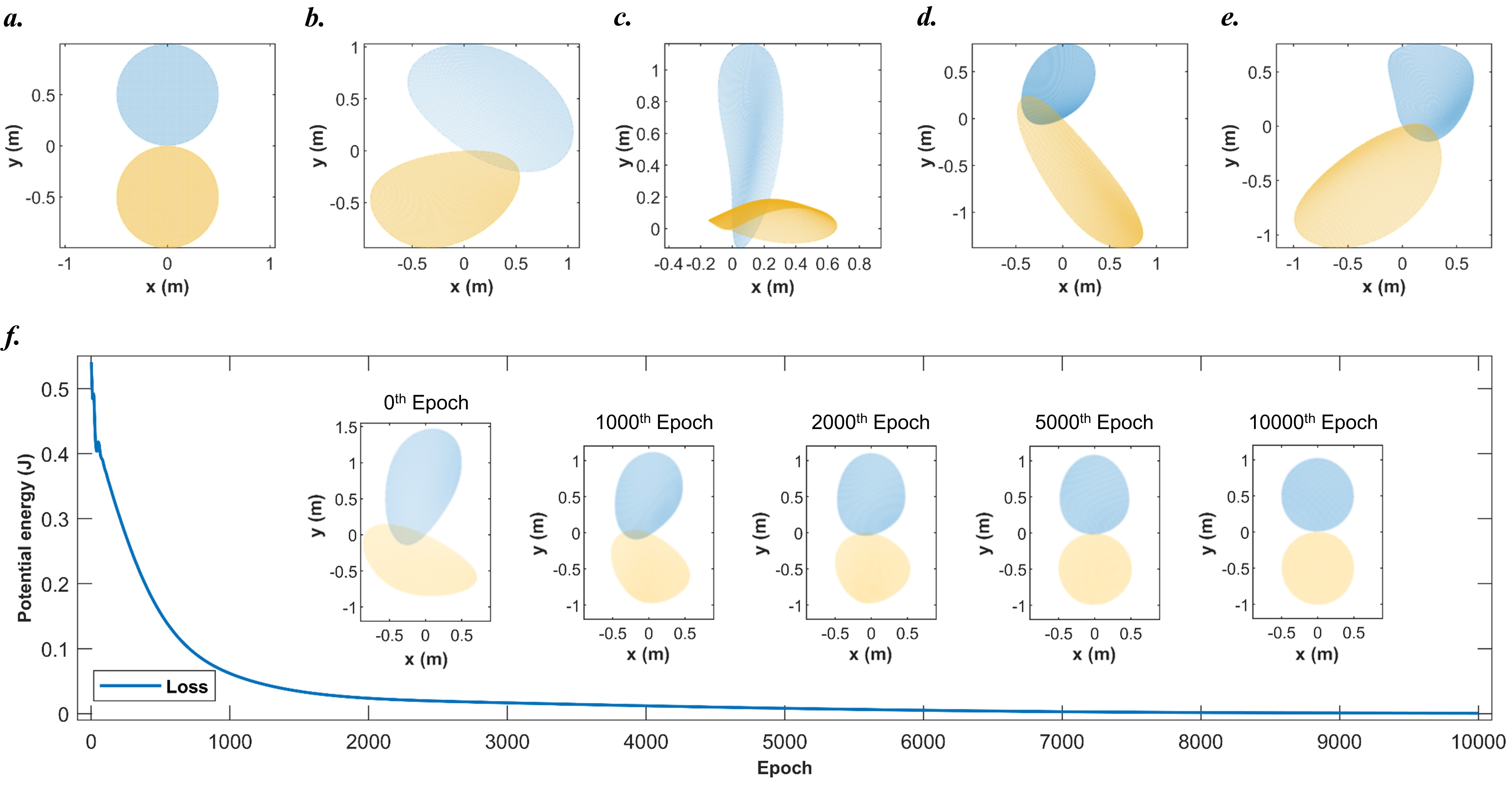}
    \caption{The multiple bodies overlapping issue caused by neural network initialisations and the relaxation scheme.  (a) Consider two circles (both radius is 0.5 m) placed at $(0,0.5)$ and $(0,-0.5)$. The displacement fields $u$ and $v$ of each circle are predicted by two individual neural networks.  (b-e) The initial configurations of circles predicted by neural networks after initialisation without training. Overlapping between two circles can be observed. (f) The loss histogram of two circles by the relaxation scheme. Note that the ADAM optimiser is used with a learning rate of $1\times10^{-4}$.}
    \label{fig:overlap_ill}
\end{figure}

After initialising neural networks, the displacement mappings were established. However, those randomly built-up mappings may cause the multiple bodies to overlap with each other, as the example shown in \Cref{fig:overlap_ill} (a-e).  Therefore, to avoid the overlapping issue, relaxation can be applied first before solving problems. In relaxation, the contact potential is not activated in the loss function. In this manner, the neural network displacement mappings will be trained to dissipate the overall potential energy, and eventually reach their undeformed states, as shown in \Cref{fig:overlap_ill} (f).

As mentioned in \Cref{sec:2:3}, neural network training has a pseudo-dynamics nature. If the increment of one training epoch goes too far, the contact bodies may penetrate each other. To address this problem, the learning rate of training should be properly selected based on the pseudo-velocity assessed by \Cref{eq:pseudo-veloctiy}. Moreover, since the soft boundary condition imposition technique is applied for the moving boundary conditions, a gradual loading formula can be applied to circumvent the body penetration
\begin{equation}
    \Pi_{\text{EBC}}(t)=\frac{1}{n} \sum \Vert\boldsymbol{u}(\boldsymbol{x})-\frac{t}{t_{\max}}\bar{\boldsymbol{u}}(\boldsymbol{x})\Vert_{2}^{2},
\end{equation}
where $t$ is the current training epoch and $t_{\max}$ is the maximum training epoch, $n$ is the number of sample points on the essential boundary. As such, the soft boundary condition is gradually imposed through the training loop. We also highlight that, when the displacement boundary condition is roughly fulfilled with a small residual and the contact system is stabilised, the hard boundary imposition techniques can be then utilised instead. 

In general, the initialising schemes for neural networks are likely to confine the initial output of neural networks within the range of $[-1,1]$. Furthermore, given that bias is typically not assigned to the final hidden layer in neural networks, the significant responsibility for scaling the output value is placed on the weights of this layer when solely the tanh activation function is employed. Hence, when the displacement fields only have small deformation, the training of neural networks will be very difficult and time-consuming. To address this problem, the output scaling factors are added to the established neural network mapping (\Cref{eq:Modified_NN_Map})
\begin{equation}
    u(\boldsymbol{x})=\xi (\boldsymbol{F}(\boldsymbol{x})\odot \boldsymbol{g}(\boldsymbol{x})+\bar{\boldsymbol{u}}(\boldsymbol{x})),
\end{equation}
where $\xi$ is the pre-defined scaling factor. The output scaling is used in the following benchmark problems (the Hertz contact problems).

\subsection{Benchmark tests and discussions}
\label{sec:3:4}
\begin{figure}
\centering
\includegraphics[scale=0.7]{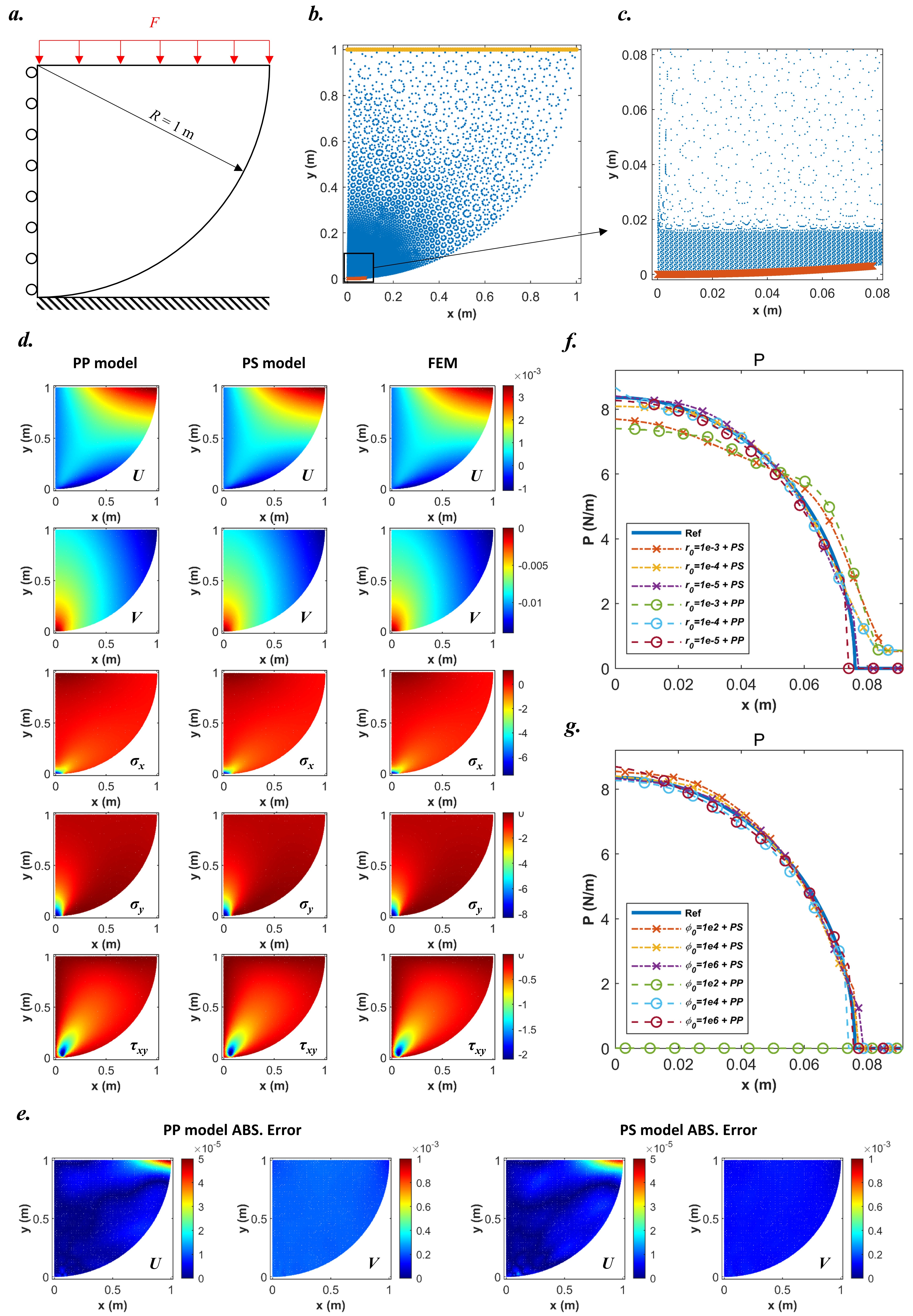}
\caption{Results for the Hertz contact problem. (a) The configuration of the Hertz contact problem; (b) The sample points distribution in the computational domain; (c) The refined sample points near the contact area. (d) Comparisons of the displacement and stress contours from the proposed PINN frameworks and the FEM. (e) The absolute error contours of the PINN frameworks. (f) Comparisons of the contact pressure plots with different $r_0$; (g) Comparisons of the contact pressure plots with different $\phi_0$. PS and PP refer to point-to-surface and point-to-point.}
\label{fig:Hertz_all_results}
\end{figure}

Using the aforementioned framework, the 2D Hertz contact problem is analysed. The problem is under the assumption of small deformation. A half cylinder is compressed on a rigid horizontal plane, as shown in \Cref{fig:Hertz_all_results} (a). A uniform pressure of $F = 0.5 $ N/m is subjected to be top of the half cylinder. A linear elastic material is considered, where Young’s modulus $E = 200 $ Pa and Poisson’s ratio $\nu = 0.3$. The analytical solution is obtained from \cite{Franke2010}. To solve this problem, two FNNs are built up to predict the displacements $U$ and $V$, respectively. Each network contains 3 hidden layers and 30 neurons per layer. The scaling factor is $\xi = 1\times10^{-3}$. Thus, the $U$ and $V$ outputs of neural networks are obtained by
\begin{equation}
    \begin{aligned}
        U&=\xi xF_{x}(x,y),\\
        V&=\xi F_{y}(x,y),
    \end{aligned}
\end{equation}
where the hard boundary condition is applied to the $U$ prediction.

The sample points distribution is shown in \Cref{fig:Hertz_all_results} (b). In this work, the sample points inside the computational domain are generated following the Gaussian quadrature. Since the contact area is located at the bottom of the cylinder, refined sample points are applied near the contact area, as shown in \Cref{fig:Hertz_all_results}(c). The ADAM optimiser is selected with a learning rate of $1\times10^{-4}$. For each case, neural networks are trained 5 times and each training converges by $2\times10^3$ epochs. Both the point-plane and point-point contact models are tested. 

\Cref{fig:Hertz_all_results}(d) shows the results produced by the PINN-based framework and the FEM. \Cref{fig:Hertz_all_results}(e) further shows the absolute error contours of displacement fields of the proposed PINN frameworks. The mean absolute error of the predicted $U$ and $V$ fields by using the PP model are respectively $3.89\times10^{-6}$ N/m$^2$ and $2.15\times10^{-4}$ N/m$^2$, while those of the PS model are respectively $4.36\times10^{-6}$ N/m$^2$ and $1.28\times10^{-4}$ N/m$^2$. All predicted contours are in good agreement with the reference results. \Cref{fig:Hertz_all_results} (f) and (g) plots the contact pressure obtained by the PINN-based framework and \Cref{tab:table1} lists the mean relative percentage errors of PINN predictions with different contact models.  The discussion starts with using different $r_0$ while a fixed $\phi_0$ of $1\times10^4$ is applied. We note that $r_0$ here also controls the spacing of the contact sample points on the contact area, in other words, a decreasing $r_0$ will induce more contact points on the same length of a contact boundary. With decreasing selections of $r_0$, the predicted contact pressure fits the analytical solution better, especially near the area of the transition point from contact to the free surface, as shown in \Cref{fig:Hertz_all_results}(e).  It is rational since it should converge to the analytical solution with decreasing selection of $r_0$, analogous to the real atom model used in molecular dynamics. When changing the value of $\phi_0$ and fixing $r_0 = 1\times10^{-5}$ m, it is found that the accuracy of the results is not very sensitive to different $\phi_0$, as shown in \Cref{fig:Hertz_all_results}(f). It is also worth noting that, when using the PP contact model, the cylinder can penetrate the substrate plane if too small $\phi_0$ is selected. Notably, $\phi_0$ is similar to the penalty factors in the penalty method when imposing the essential boundary conditions. Moreover, it can be found that the framework with the PS model can achieve greater accuracy. Despite the relatively large discrepancy, the framework with the PP contact model is still effective enough to accurately capture the stress concentration patterns near the contact area, as presented in \Cref{fig:Hertz_all_results}(d). Therefore, for irregular contact surfaces, the PP contact model is used instead of the PS contact model. 

Finally, \Cref{tab:table2} presents the computational accuracy and efficiency of the energy-based PINNs with respect to different neural network sizes. With increasing numbers of layers and neurons per layer, neural networks can provide more accurate contact pressure predictions. However, the computational efficiency decreases due to the increasing number of trainable parameters in larger networks. Notably, training networks of the $3\times20$ size costs more time than larger networks. This is because a GPU is used. The GPU spends more time in accessing data rather than calculations. The computational efficiency should be increased with decreasing size of networks if a CPU is used.

\begin{table}
\begin{centering}
\caption{The mean absolute percentage error of PINN predictions by using different $r_0$, $\phi_0$ and contact models. PS and PP refer to point-to-surface and point-to-point, respectively. For modelling with varying $r_0$ and $\phi_0$, $\phi_0$ and $r_0$ are fixed to $1\times10^4$ and $1\times10^{-5}$, respectively.}
\label{tab:table1}
\begin{tabular}{cccccccc}
\toprule 
\multirow{2}{*}{Contact Model} & \multicolumn{3}{c}{$r_{0}$} & & \multicolumn{3}{c}{$\phi_{0}$}\tabularnewline
\cmidrule{2-4} \cmidrule{6-8}
 & $1\times10^{-3}$ & $1\times10^{-4}$  & $1\times10^{-5}$ &  & $1\times10^2$ & $1\times10^4$ & $1\times10^6$ \tabularnewline
\midrule
PS &  15.40\%&  4.40\%&  3.87\%&  & 5.11\%&  3.87\%& 3.39\% \tabularnewline
PP &  18.31\%&  11.71\%&  6.83\%  & & $\backslash$ &  6.83\%& 4.98\% \tabularnewline
\bottomrule
\end{tabular}
\par\end{centering}
\end{table}

\begin{table}
\begin{centering}
\caption{The relative mean absolute error (RMAE) and GPU training time of the contact pressure predicted by the energy-based PINNs using different neural network sizes. The PP model is selected. $\phi_0$ and $r_0$ are $1\times10^4$ and $1\times10^{-5}$, respectively. For each case, neural networks are trained 5 times and each training converges by $5\times10^3$ epochs. The network size is presented by the number of hidden layers times the number of neurons per layer.}
\label{tab:table2}
\begin{tabular}{c|cccc|cc}
\toprule
Network Size & RMAE P & GPU Time (s) & ~ & Network Size & RMAE P & GPU Time (s) \\ 
\cmidrule{1-3} \cmidrule{5-7} 
$2\times30$ & 4.46\% & 159.40 & ~ & $3\times20$ & 7.58\% & 222.62  \tabularnewline
$3\times30$ & 4.02\% & 204.55 & ~ & $3\times30$ & 4.02\% & 204.55 \tabularnewline
$4\times30$ & 2.82\%  & 261.46 & ~ & $3\times40$ & 3.20\% & 210.47 \tabularnewline 
\bottomrule
\end{tabular}
\par\end{centering}
\end{table}

\section{Numerical examples}
\label{sec:sec4}
Herein, three frictionless contact problems with material and geometric nonlinearities are conducted to test the performance of the proposed framework. In all cases, the hyperelastic Neo-Hookean material is applied. Relaxation is also applied before all loading processes. The plain strain condition is considered. The neural network is built up by the TensorFlow 2 library in Python. The ADAM optimiser is selected with a learning rate of $1\times10^{-5}$. All the problems are modelled on a $12^{\text{th}}$ Gen Intel(R) Core(TM) i5-12490F CPU (3.0 GHz) and with a Geforce 4060 Ti GPU. 

\subsection{Rubber ironing}
\begin{figure}
    \centering
    \includegraphics[width=0.9\linewidth]{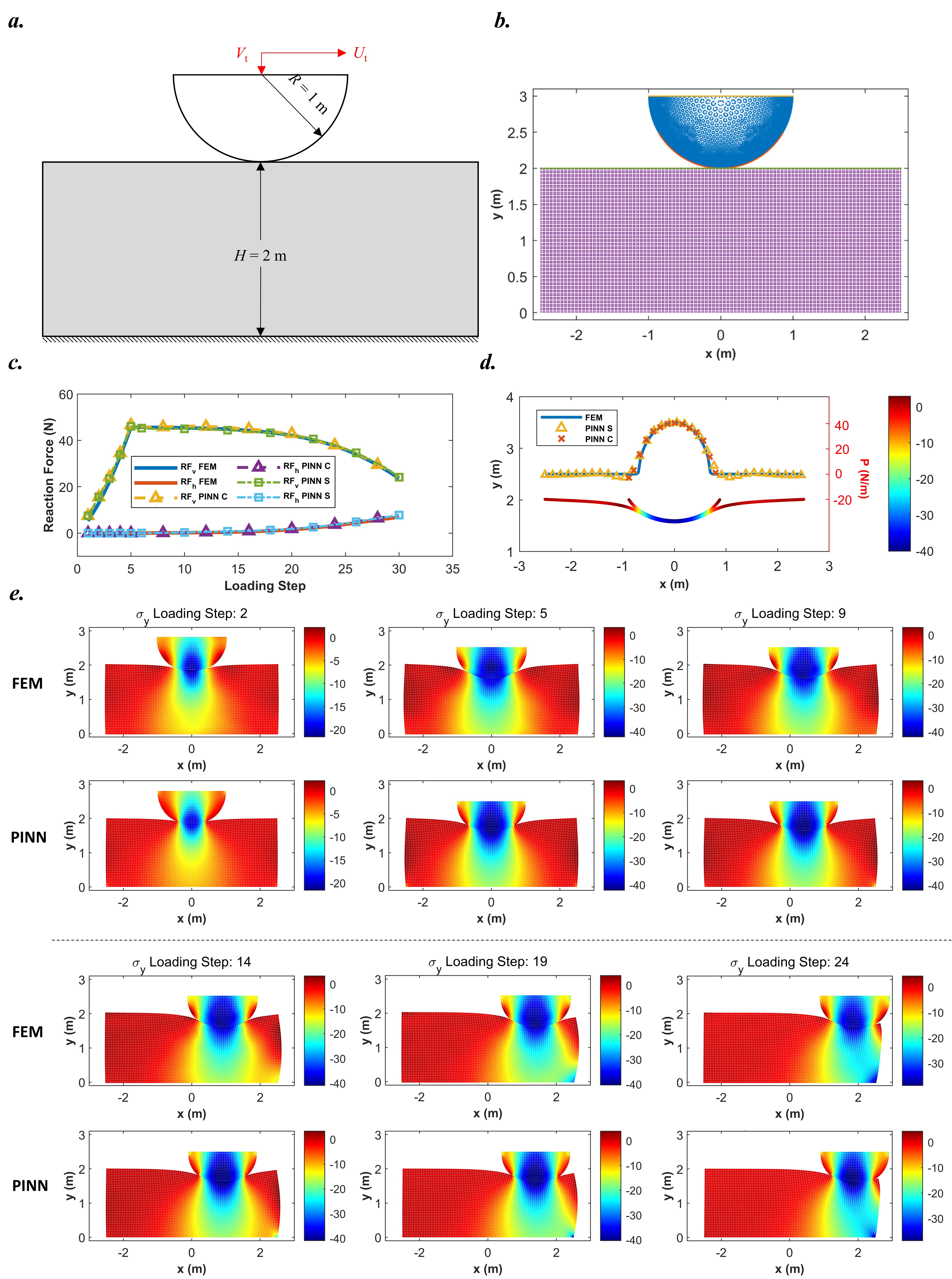}
    \caption{The rubber ironing problem. (a) Configuration of the rubber ironing problem; (b) The sample point distribution in the computational domain; (c) Vertical and horizontal reaction forces (RF) during the loading process. C and S mean that the RFs are obtained from the boundary on the half cylinder and the slab; (d) The contact pressure plot and the contour at the $5^{\text{th}}$ loading step; (e) Comparisons of $\sigma_{y}$ stress contours from the proposed PINN framework at different loading steps.}
    \label{fig:Ironing_all_results}
\end{figure}

The ironing problem is one of the most prevailing benchmarks for frictionless contact problems under large deformation. The configuration of the problem is presented in \Cref{fig:Ironing_all_results} (a), where a half cylinder is first compressed vertically on a slab and then slided horizontally. The sample points distribution for this problem is shown in \Cref{fig:Ironing_all_results} (b). Throughout the whole modelling, the bottom boundary of the slab is fixed. For the half cylinder, during compressing, a vertical displacement of $V_{\text{t}}=-0.5$ m is applied by five uniform loading steps, while the horizontal displacement is fixed. During sliding, a horizontal displacement of $U_{\text{t}}=2.5$ m is achieved by 25 uniform loading steps, while the compressing is remained.  The Young’s modulus for the half cylinder rubber and the slab are $3\times10^2$ Pa and $1\times10^2$ Pa, respectively, and their Poisson ratios are $0.3$. Four FNNs are established with 3 hidden layers and 30 neurons per layer. To impose the displacement boundary conditions, the outputs of neural networks are constructed by
\begin{equation}
    \begin{aligned}
        U_{\text{c}}&= (y-3)u_{\text{c}}(x,y)+U_t,\\
        V_{\text{c}}&= (y-3)v_{\text{c}}(x,y)+V_t,\\
        U_{\text{s}}&= yu_{\text{s}}(x,y),\\
        V_{\text{s}}&= yv_{\text{s}}(x,y).
    \end{aligned}
\end{equation}
We note that each loading step is trained by $10$ training sessions and $2\times10^3$ epochs per session. The contact sample points on the half cylinder and the slab are placed with a spacing of $r_0=1\times 10^{-2}$ m. The pre-defined potential constant $\phi_0 = 1\times 10^2$. The penalty factor is $\kappa = 1\times 10^4$.

\Cref{fig:Ironing_all_results} (c) plots the vertical and horizontal reaction forces (RFs) during the whole loading process. Note that C and S in the legend refer to the RFs calculated on the upper boundary of the half cylinder and the bottom boundary of the slab. In the previous five loading steps, the vertical RF gradually increases due to the compression. The horizontal RFs remain zero due to the problem's symmetry. In the following loading steps, because the symmetry of the problem is broken, the horizontal RF slowly increases, while the vertical RF decreases. The contact pressure distribution and contour at the $5^{\text{th}}$ loading step are given in  \Cref{fig:Ironing_all_results} (d). Both the contact pressure results obtained from the slab and cylinder align well with the FEM results. Moreover, compared to FEM results, the relative mean absolute differences of the contact pressure from the half cylinder and the slab are $6.81\%$ and $6.41\%$, respectively. The $\sigma_y$ stress contours at six selected loading steps are presented in  \Cref{fig:Ironing_all_results} (e).  The FEM results are also presented for comparison. It can be observed that not only the RF plots obtained by the proposed PINN agree well with the FEM results, but the stress contours predicted by the PINN align closely to the FEM contours as well. Therefore, it is concluded that the frictionless condition is achieved during the whole modelling. 

However, slight departures can still be found when comparing the stress contours, especially near the edges of the contact area at the previous 5 loading steps, as shown in \Cref{fig:Ironing_all_results} (c) and (e). This is because the spacing of the contact sample points on the contact surfaces is still too large. As discussed in \Cref{sec:3:4}, the computational accuracy around the contact area can be further improved by decreasing the spacing of contact sample points.   

\subsection{Rubber ring contact instability}
\begin{figure}
    \centering
    \includegraphics[width=1\linewidth]{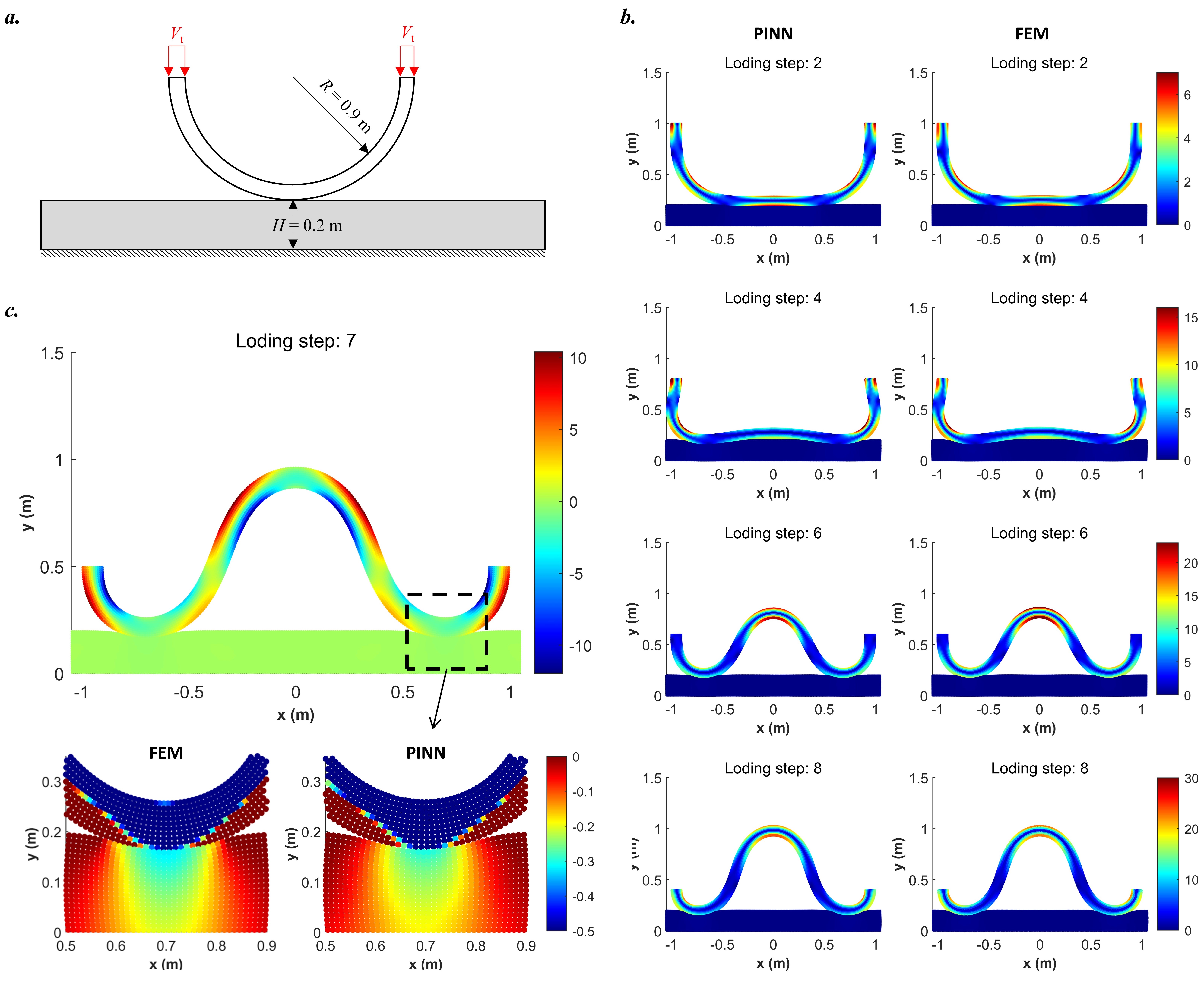}
    \caption{The rubber ring contact instability problem. (a) Configuration of the ring and the slab; (b) Comparisons of the von Mises stress contours from the proposed PINN framework and the FEM at different loading steps; (c) $\sigma_{y}$ stress contour obtained by the FEM and the proposed PINN at the $7^{\text{th}}$ loading step.}
    \label{fig:Ring_ironing}
\end{figure}

Consider a half rubber ring compressed on the rubber slab, as shown in \Cref{fig:Ring_ironing} (a). The Young’s modulus of the ring and the slab are 100 Pa and 1 Pa, respectively. The Poisson's ratios of the ring and slab are 0.3. The thickness of the ring is 0.1 m. The displacement boundary condition, $V_{\text{t}} = 0.8$ m, is imposed on the rings, and the loading is done by 8 uniform loading steps. Relaxation is applied. The plain strain condition is considered. Four FNNs are established with 3 hidden layers and 50 neurons per layer. To impose the displacement boundary conditions, the outputs of neural networks are constructed by
\begin{equation}
    \begin{aligned}
        U_{\text{r}}&= (y-1.2)u_{\text{r}}(x,y),\\
        V_{\text{r}}&= (y-1.2)v_{\text{r}}(x,y)+V_t,\\
        U_{\text{s}}&= yu_{\text{s}}(x,y),\\
        V_{\text{s}}&= yv_{\text{s}}(x,y).
    \end{aligned}
\end{equation}
Each loading step is trained by $5$ training sessions and $5\times10^4$ epochs per session. The contact sample points on the half cylinder and the slab are placed with a spacing of $r_0=3\times 10^{-4}$ m. The pre-defined potential constant $\phi_0 = 1\times 10^5$. The penalty factor is $\kappa = 1\times 10^6$.
 
The predicted von Mises stress contours are presented in \Cref{fig:Ring_ironing} (b). As observed, the rubber ring gradually attaches the slab with an increasing contact area when $V_{\text{t}} = 0.4$ m. The von Mises stress concentration mainly appears near the arc surfaces due to the bending of the ring. Interestingly, when $V_{\text{t}} = 0.4$ m, the ring bends inversely and the original contact area is divided into two parts. With further compression,  the inverse bending appears more apparent. Besides, the von Mises stress concentration getting more severe at the middle of the half ring. We note that the contact of the two bodies also compresses the rubber slab, although it is not obvious in the stress contours. \Cref{fig:Ring_ironing} (c) further shows the $\sigma_y$ contour at the $7^{\text{th}}$ loading step. Moreover, the contact area contour is compared with the FEM result. The contours from the proposed PINN framework are in good agreement with those from the FEM and only small differences can be found. 

\subsection{Compression of two rubber rings}
\begin{figure}
    \centering
    \includegraphics[width=1\linewidth]{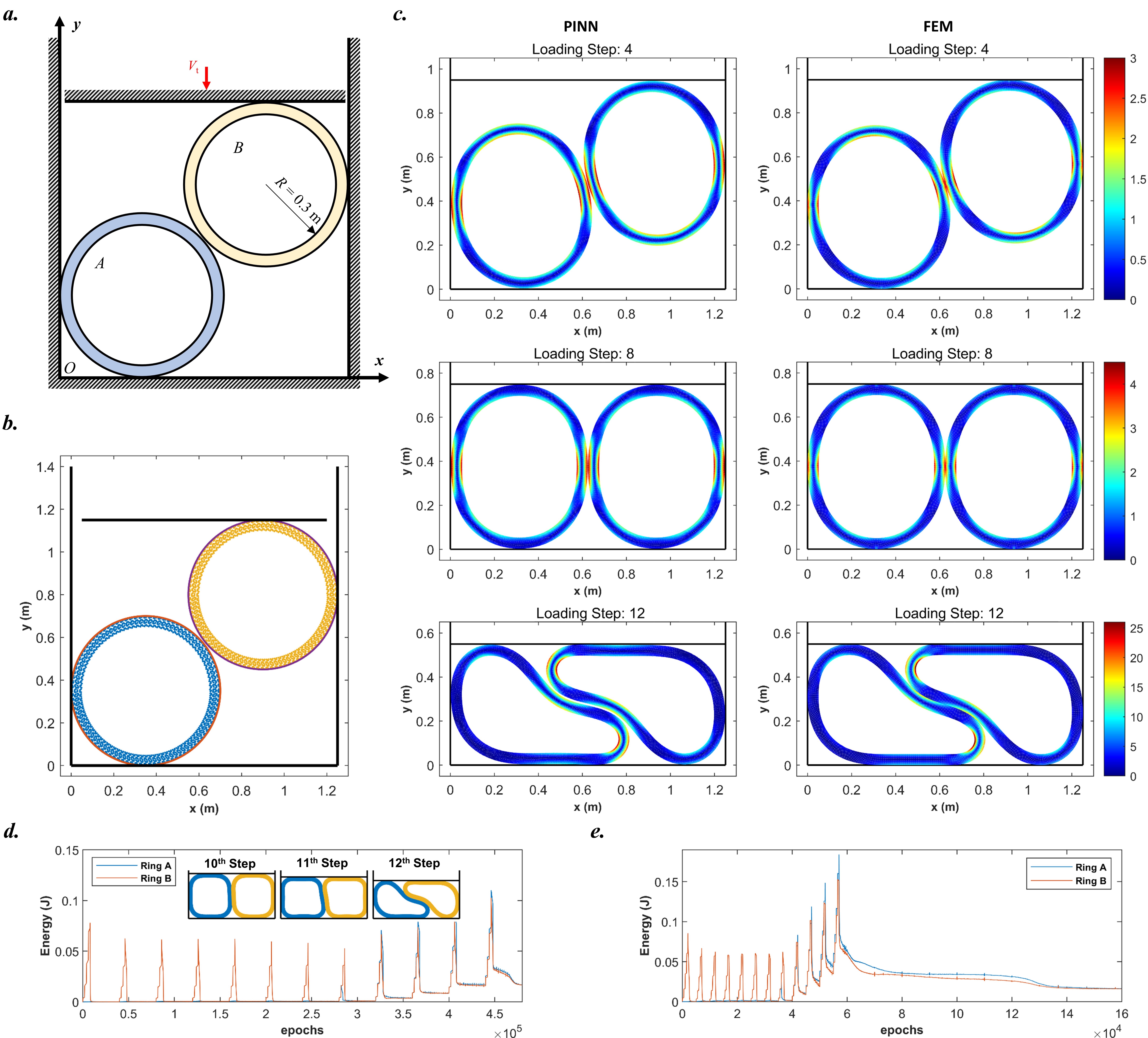}
    \caption{The two rubber rings compression. (a) Configuration of the two rings, the rigid sink and the rigid cap; (b) The initialised sample points in the computational domain. (c) Comparisons of the von Mises stress contours from the proposed PINN framework and the FEM at different loading steps; (d) The potential energy of the two rings during the whole loading process (each loading step is trained 40 times and $2\times10^3$ epochs per training); (e) The potential energy of the two rings during the whole loading process. In this case, each loading step can be first trained $10$ times and $5\times10^2$ epochs per training. After twelve-step loading, an additional 10 training sessions with $1\times10^4$ epochs per training are conducted to fine-tune the neural networks.}
    \label{fig:rings_all_results}
\end{figure}

In this problem, two rubber rings are placed in a sink and compressed by a horizontal cap, as shown in \Cref{fig:rings_all_results}(a). The Young’s modulus and Poisson's ratio of the two rings are identically 100 Pa and 0.3, respectively. The inner radius and thickness of the rings are 0.3 m and 0.05 m, respectively. The centres of the two rings are located at $(0.35,0.35)$ and $(0.9,0.8)$. A vertical displacement, $V_{\text{t}}=-0.6$ m, is applied on the horizontal cap, compressing the two rubber rings to deform downward. The loading is done by 12 uniform loading steps. Four FNNs are established with 3 hidden layers and 50 neurons per layer. Since no explicit displacement boundary condition is required for the rings, the vanilla FNN structure is directly used. The PP model is applied for the ring-ring contact, while the PS contact model is applied for ring-sink and ring-cap contacts. Each loading step is trained 20 times and $2\times10^3$ epochs per training. The contact sample points on the rings are identically placed with a spacing of $r_0=1\times 10^{-3}$ m, as shown in \Cref{fig:rings_all_results} (b). The pre-defined potential constants for the PP model and the PS model are  $\phi_0 = 1\times 10^{-2}\ \text{and}\ 1\times 10^{1}$, respectively.  

\Cref{fig:rings_all_results}(c) shows the von Mises stress contour of the compressed rings at different loading steps. Results obtained from the ABAQUS are also presented for comparison. It is worth highlighting that the automatic stabilization with a specific damping factor of $2\times10^{-4}$ must be applied when using ABAQUS. This is not implemented in the proposed PINN framework. This additional robustness is contributed by the use of the ADAM optimiser. The Adam optimiser harnesses adaptive learning rates for each individual trainable parameter by assessing both the first and the second moments of loss gradients. Besides, incorporating momentum (i.e., maintaining an exponentially weighted moving average of past loss gradients) smoothens the optimisation process and avoids loss oscillations \cite{kingma2014adam}. As observed, with the rigid cap moving downward, the upper ring is compressed and slides into the vacancy between the bottom ring and the sink. Those two rings squeeze each other and hold an equilibrium state. With increasing compression, the symmetry equilibrium condition is broken down. Due to the frictionless condition, rings exhibit chirality extrusion deformation. Such instability is captured in the loss histogram, as depicted in \Cref{fig:rings_all_results} (d). Along with the loading processes, the stored strain energies of the two rings increase dramatically at the beginning of each loading step. For example, in the $12^{\text{th}}$ loading step (started after $4.4\times10^{5}$ epochs), the stored strain energy of a single rubber ring increases to 0.12 J within the first $8\times10^{3}$ epochs. During the following training in a loading step, the strain energy gradually dissipates. It is obvious that the stored strain energy of the chirality configuration is lower than that of the symmetrical configuration, posing that the chirality deformation is the result of an instability phenomenon. 

It is highlighted that, the modelling process is simpler and more straightforward while using the proposed PINN framework. Furthermore, once the neural network is well-trained, results in terms of displacements and stresses at any position can be directly exported by feeding the coordinate as an input. If only the final configuration of the compressed rings is required, it is not necessary to train the neural networks to converge at every loading step. For example, each loading step can be first trained 10 times and $5\times10^2$ epochs per training. After twelve-step loading, an additional 10 times training with $1\times10^4$ epochs per training are conducted to fine tune the neural networks. As the loss histogram shown in \Cref{fig:rings_all_results} (e), the whole solving process can also stably converge after $1.4\times10^5$ epochs.  In this manner, the CPU time of the proposed PINN framework for this compression case is 37 mins (3767 sample points for each ring), while the ABAQUS costs 30 mins (3468 nodes for each ring). Currently, our implementation leverages the GPU functionality inherently integrated within the TensorFlow framework. The computational efficiency of the proposed energy-based PINNs could be enhanced through code optimization, which will be addressed in future research endeavours.

\section{Conclusions}
\label{sec:sec5}
In this work, we proposed an energy-based PINN framework for solving contact problems under large deformations. Thanks to the natural pseudo-dynamics of PINN training, the neural network training can be treated as a loading process, which ensures the framework's robustness. An exponential surface contact potential is applied to mimic the microscopic LJ potential, which intrinsically prevents solid bodies from penetrating each other. Besides, numerical schemes including relaxation, gradual loading and output scaling are proposed, further improving the stability and accuracy of the proposed framework for solving contact problems. The well-known Hertz contact benchmark problem is conducted, demonstrating the effectiveness of the proposed framework. Then, complex contact problems including the material nonlinearity as well as the large deformation are tested by using the proposed PINN framework. 

It is worth highlighting that the proposed PINN framework offers a very easy way to model contact problems. Physics in terms of the energy functional, penalty function of essential BC and contact surface potential are straightforwardly integrated into the physics-informed loss function.  It is very robust with respect to nonlinear problems. Besides, in the solving process, only one neural network training loop is required, while different iterative methods are needed to deal with different nonlinearities in the traditional computation methods. Moreover, as demonstrated by numerical examples, the energy-based PINN framework can easily capture the instability phenomenon without adding any artificial perturbation. In addition, for complex contact problems, the PINN framework can provide compatible computational efficiency when compared with commercial FEM software. 

While the proposed Physics-Informed Neural Network (PINN) framework demonstrates excellent performance in addressing complex contact problems, it is important to acknowledge its limitations in frictionless scenarios, necessitating the consideration of frictional contact. The contact detection algorithms can also be integrated into the energy-based PINNs to alleviate the training expense. Besides, it should be noted that, due to the use of neural networks (i.e., nonlinear computing systems), neither the computational efficiency nor the convergence rate of the energy-based PINNs is comparable with traditional methods when dealing with simple solid mechanics problems. Hence, more attention should be paid to the complex mechanics problems that are very hard or even prohibitive for traditional methods. In addition, despite the competitive computational efficiency, the great potential of deep learning for scientific modelling has yet to be fully unleashed. Such physics-informed energy-based loss function can be integrated into more advanced deep learning models to further improve the computational efficiency, for example, the operator learning techniques \cite{Lu2021,li2020fourier,wang2024homogenius}.  These will be covered in our future works.

\section*{Author contribution}
\textbf{J. Bai:} Conceptualisation, Methodology, Coding, Formal analysis, Writing-Original draft. \textbf{Z. Lin:} Methodology, Coding, Formal analysis, Writing-review \& editing. \textbf{Y. Wang:} Methodology, Formal analysis, Writing-review \& editing. \textbf{J. Wen:} Formal analysis, Writing-review \& editing. \textbf{Y. Liu:} Writing-review \& editing. \textbf{T. Rabczuk:} Formal analysis, Writing-review \& editing. \textbf{Y.T. Gu:} Conceptualisation, Methodology, Formal analysis, Writing-review \& editing. \textbf{X.-Q. Feng:} Supervision, Writing-review \& editing, Funding acquisition.

\section*{Declaration of competing interest}
The authors declare no competing interests.

\section*{Acknowledgments}
Support from the National Natural Science Foundation of China (Grant nos. 11921002, 12032014 and T2488101) is acknowledged (J. Bai and X.-Q. Feng). Support from the Australian Research Council research grants (IC190100020 and DP200102546) is also acknowledged (J. Bai and Y. Gu).

\appendix
\setcounter{figure}{0}
\section{The pseudo-velocity of the energy-based PINNs during the training process}
\label{sec:sample:appendix}
Recall \Cref{eq:Neural_net_mapping}, the displacement prediction before the $(t+1)^{\text{th}}$ epoch is calculated by
\begin{equation}
    u^{t}=F(\boldsymbol{x};\boldsymbol{\theta ^ {\text{t}}}).
\end{equation}
\noindent Here, the superscript denotes the number of the current training epoch. After the  $(t+1)^{\text{th}}$ epoch through gradient descendant optimisers, $\boldsymbol{\theta }^{t}$ is updated by
\begin{equation}
    \boldsymbol{\theta }^{t+1}=\boldsymbol{\theta }^{t}-\eta \frac{\partial \Pi}{\partial \boldsymbol{\theta}}|_{\boldsymbol{\theta }^{t}}.
\end{equation}
\noindent Therefore, the increment of $\boldsymbol{\theta}$ can be obtained as
\begin{equation}
    \Delta \boldsymbol{\theta}^{t+1}=\boldsymbol{\theta}^{t+1}-\boldsymbol{\theta}^{t}=-\eta \frac{\partial \Pi}{\partial \boldsymbol{\theta}}|_{\boldsymbol{\theta }^{t}}.
    \label{eq:gradient_flow}
\end{equation}
Then, the Taylor expansion is performed
\begin{equation}
\begin{aligned}
      u^{t+1}(\boldsymbol{x};\boldsymbol{\theta}^{t+1}) &= F(\boldsymbol{x};\boldsymbol{\theta}^{t})+\frac{\partial F(\boldsymbol{x};\boldsymbol{\theta}^{t})^{\text{T}}}{\partial \boldsymbol{\theta}}(\Delta \boldsymbol{\theta}^{t+1})+O(\Delta\boldsymbol{\theta}^{t+1}) \\
      &\approx F(\boldsymbol{x};\boldsymbol{\theta}^{t})+\frac{\partial F(\boldsymbol{x};\boldsymbol{\theta}^{t})^{\text{T}}}{\partial \boldsymbol{\theta}}(\Delta \boldsymbol{\theta}^{t+1}),
\end{aligned}
\end{equation}
\noindent where $O(\Delta\boldsymbol{\theta}^{t+1})$ represents the higher order small quantities in the expansion. Thus, the increment of displacement after one epoch of training gives
\begin{equation}
    \Delta u^{t+1}=u^{t+1}-u^{t}\approx \frac{\partial F(\boldsymbol{x};\boldsymbol{\theta}^{t})^{\text{T}}}{\partial \boldsymbol{\theta}}(\Delta \boldsymbol{\theta}^{t+1}).
    \label{eq:du}
\end{equation}
By substituting \Cref{eq:gradient_flow} into \Cref{eq:du}, one can obtain
\begin{equation}
    \Delta u^{t+1}\approx -\eta \frac{\partial F(\boldsymbol{x};\boldsymbol{\theta}^{t})^{\text{T}}}{\partial \boldsymbol{\theta}}\frac{\partial \Pi}{\partial \boldsymbol{\theta}}.
\end{equation}
The learning rate $\eta$ controls the stepping length of displacement prediction. When the selection of $\eta$ tend to be infinitely small, the flow of $u$ with respect to the continuous training step $t$ in a continues manner, e.g., a pseudo-velocity during the training process
\begin{equation}
   \frac {\partial u}{\partial t}=\lim_{\eta \to 0} \frac{\Delta u^{t+1}}{\eta}=-\frac{\partial F(\boldsymbol{x};\boldsymbol{\theta}^{t})^{\text{T}}}{\partial \boldsymbol{\theta}}\frac{\partial \Pi}{\partial \boldsymbol{\theta}}|_{\boldsymbol{\theta }^{t}}.
\end{equation}
The same conception has also been applied to study dynamical property of $\boldsymbol{\theta}$ during training
\begin{equation}
    \frac {\partial \boldsymbol{\theta}}{\partial t}\approx- \frac{\partial \Pi}{\partial \boldsymbol{\theta}}.
\end{equation}
where $\partial \boldsymbol{\theta} / \partial t$ is the gradient flow \cite{SifanWang_Gradientflow}. 

The pseudo-velocity approach is similar to the pseudo-time stepping method in computational fluid dynamics, which specifically transforms steady-state problems into transient problems. Pseudo-time stepping is a commonly used numerical method for nonlinear steady-state problems, and we apply this concept within energy-based PINNs. Consequently, a numerical pseudo-dynamic process can be obtained in the hyperelastic computation, allowing for a better understanding of the training process of energy-based PINNs.

\section{Gradient descendant algorithms for the nonlinear cantilever beam problem}
\label{sec:app_gd_nolin}
\begin{figure}
    \centering
    \includegraphics[width=1\linewidth]{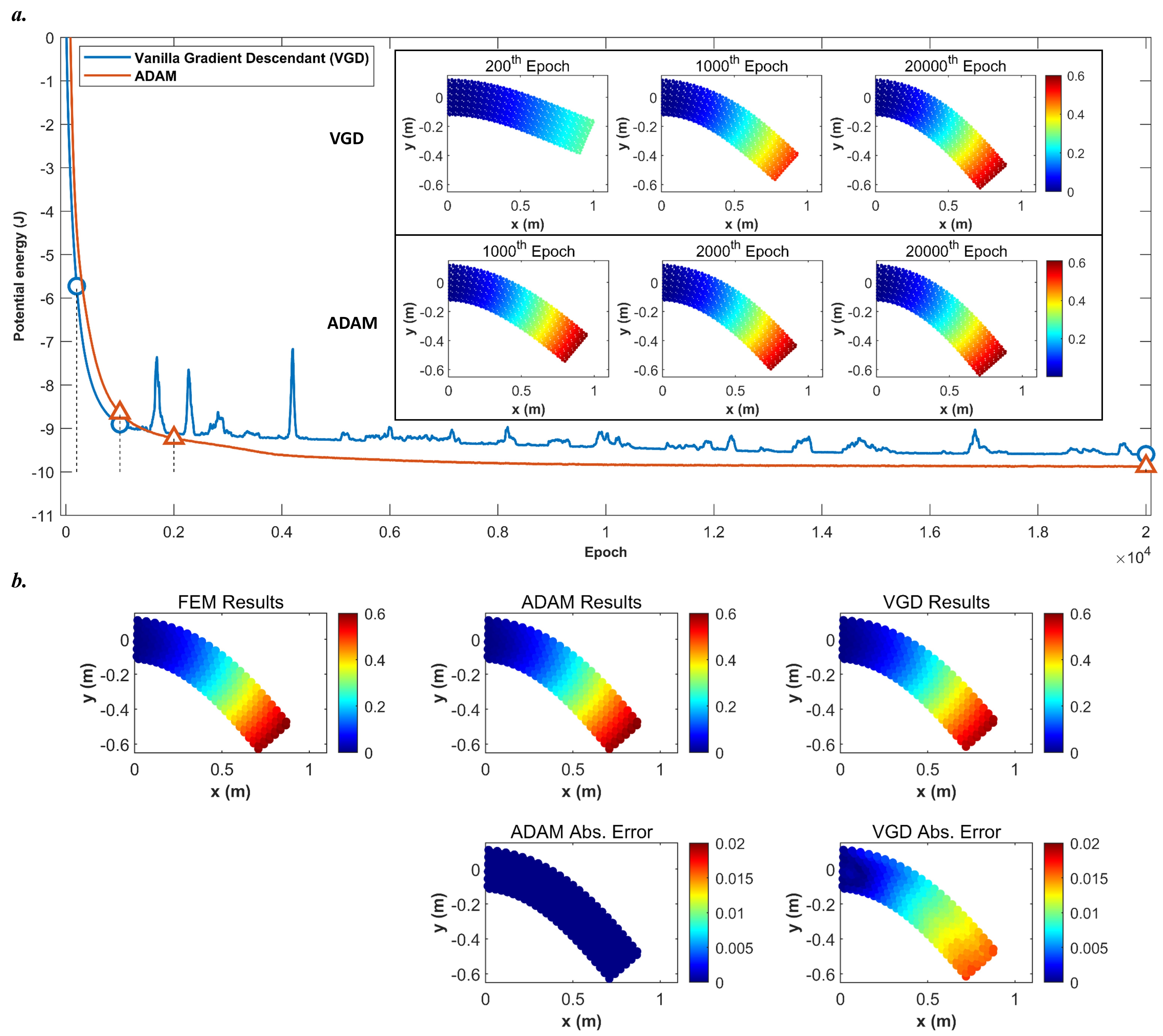}
    \caption{(a) The training dynamics of energy-based PINNs for solving a cantilever beam problem with neo-Hookean material and large deformation. A parabolic distributed force of  $30$ N is downwardly applied on the right boundary. (b) Absolute displacement contours from the FEM, the ADAM optimiser and the VGD algorithm and the corresponding absolute error contours.}
    \label{fig:app_energy_dissipation}
\end{figure}

Herein, we consider the nonlinear version of the cantilever beam problem presented in \Cref{sec:2:3}. The neo-Hookean material and geometrical nonlinearity are applied. The Young's modulus, the Poisson's ratio and the geometry of the problem are the same. Given that the problem involves nonlinearities, two larger feedforward neural networks with $3$ hidden layers and $10$ neurons per layer are used to predict the displacements. Again, both the vanilla gradient descendant algorithm and the ADAM optimiser are used to train neural networks. The results are presented in \Cref{fig:app_energy_dissipation}.  It is obvious that the training of PINNs can be regarded as a pseudo-dynamic bending process. The training by using the ADAM optimiser can stably converge to the final solution, while that by using the VGD algorithm suffers oscillations, as shown in  \Cref{fig:app_energy_dissipation} (a). More importantly, the training by using the VGD algorithm is trapped by a local optimum and converges to a relatively high energy value, resulting in relatively larger discrepancies compared to the reference (FEM) result. This is further proofed by \Cref{fig:app_energy_dissipation} (b). We also note that the training by using the VGD frequently suffers from numerical crashes caused by the training oscillations. Consequently, in this work, we apply the ADAM optimisers for all numerical examples. 

 \bibliographystyle{elsarticle-num} 
 \bibliography{ref_file}





\end{document}